\documentclass[preprint,review,12pt]{elsarticle}

\usepackage{multirow}
 \usepackage{epsfig}

\usepackage{amssymb}

 \usepackage{lineno}

\usepackage{longtable}

\journal{Astroparticle Physics}

\begin{document}

\begin{frontmatter}





\title{Detection Potential of the KM3NeT Detector for High-Energy Neutrinos from the Fermi Bubbles}


\newcommand{\tUAb}{University of Aberdeen, United Kingdom}
\newcommand{\nUAb}{1}
\newcommand{\tUAm}{University of Amsterdam, the Netherlands}
\newcommand{\nUAm}{2}
\newcommand{\tAPC}{APC -- AstroParticule et Cosmologie -- UMR 7164 
(CNRS, Universit\'e Paris 7, CEA, Observatoire de Paris), Paris, France}
\newcommand{\nAPC}{3}
\newcommand{\tUAt}{University of Athens, Greece}
\newcommand{\nUAt}{4}
\newcommand{\tBar}{University of Barcelona, Spain}
\newcommand{\nBar}{5}
\newcommand{\tCIR}{CEA, IRFU, Centre de  Saclay, 91191 Gif-sur-Yvette, France}
\newcommand{\nCIR}{6}
\newcommand{\tCLS}{CEA-CNRS-UVSQ, LSCE/IPSL, 91198 Gif-sur-Yvette, France}
\newcommand{\nCLS}{7}
\newcommand{\tCNR}{CNR-ISMAR, La Spezia, Trieste, Genova, Italy}
\newcommand{\nCNR}{8}
\newcommand{\tMar}{CPPM, Aix-Marseille Universit\'e, CNRS/IN2P3, Marseille, France}
\newcommand{\nMar}{9}
\newcommand{\tCyp}{University of Cyprus}
\newcommand{\nCyp}{10}
\newcommand{\tDub}{Dublin Institute for Advanced Studies (DIAS), Ireland}
\newcommand{\nDub}{11}
\newcommand{\tUEr}{Erlangen Centre for Astroparticle Physics (ECAP),\newline
University of Erlangen--Nuremberg, Germany}
\newcommand{\nUEr}{12}
\newcommand{\tUHA}{Groupe de Recherche en Physique des Hautes Energies\newline 
(GRPHE)/EA3438/Universit\'e de Haute Alsace, Colmar, France}
\newcommand{\nUHA}{13}
\newcommand{\tHCM}{Hellenic Centre for Marine Research (HCMR), Greece}
\newcommand{\nHCM}{14}
\newcommand{\tHOU}{Hellenic Open University, Patras, Greece}
\newcommand{\nHOU}{15}
\newcommand{\tIFI}{IFIC -- Instituto de F\'isica Corpuscular, CSIC and 
Universitat de Val\`encia, Spain}
\newcommand{\nIFI}{16}
\newcommand{\tIFR}{IFREMER, France}
\newcommand{\nIFR}{17}
\newcommand{\tIBa}{INFN Sezione di Bari and University of Bari, Italy}
\newcommand{\nIBa}{18}
\newcommand{\tIBo}{INFN Sezione di Bologna and University of Bologna, Italy}
\newcommand{\nIBo}{19}
\newcommand{\tICa}{INFN Sezione di Cagliari and University of Sassari, Italy}
\newcommand{\nICa}{20}
\newcommand{\tICt}{INFN Sezione di Catania, Italy}
\newcommand{\nICt}{21}
\newcommand{\tUCt}{INFN Sezione di Catania and University of Catania, Italy}
\newcommand{\nUCt}{22}
\newcommand{\tIGe}{INFN Sezione di Genova and University of Genova, Italy}
\newcommand{\nIGe}{23}
\newcommand{\tLNS}{INFN Laboratori Nazionali del Sud, Catania, Italy}
\newcommand{\nLNS}{24}
\newcommand{\tLNF}{INFN Laboratori Nazionali di Frascati, Italy}
\newcommand{\nLNF}{25}
\newcommand{\tINa}{INFN Sezione di Napoli, Italy}
\newcommand{\nINa}{26}
\newcommand{\tUNa}{INFN Sezione di Napoli and University of Napoli, Italy}
\newcommand{\nUNa}{27}
\newcommand{\tIPi}{INFN Sezione di Pisa and University of Pisa, Italy}
\newcommand{\nIPi}{28}
\newcommand{\tIRo}{INFN Sezione di Roma, Italy}
\newcommand{\nIRo}{29}
\newcommand{\tURo}{INFN Sezione di Roma and University of Roma 1 ``La Sapienza'', Italy}
\newcommand{\nURo}{30}
\newcommand{\tING}{Istituto Nazionale di Geofisica e Vulcanologia (INGV), Italy}
\newcommand{\nING}{31}
\newcommand{\tUSt}{University of Strasbourg and Institut Pluridisciplinaire\newline
Hubert Curien/IN2P3/CNRS, Strasbourg, France}
\newcommand{\nUSt}{32}
\newcommand{\tISS}{Institute of Space Science, M\u{a}gurele-Bucharest, Romania}
\newcommand{\nISS}{33}
\newcommand{\tUKi}{University of Kiel, Germany}
\newcommand{\nUKi}{34}
\newcommand{\tKVI}{KVI, University of Groningen, the Netherlands}
\newcommand{\nKVI}{35}
\newcommand{\tULe}{University of Leeds, United Kingdom}
\newcommand{\nULe}{36}
\newcommand{\tULi}{University of Liverpool, United Kingdom}
\newcommand{\nULi}{37}
\newcommand{\tNCS}{National Center of Scientific Research ``Demokritos'', Athens, Greece}
\newcommand{\nNCS}{38}
\newcommand{\tNik}{Nikhef, Amsterdam, the Netherlands}
\newcommand{\nNik}{39}
\newcommand{\tNIO}{Koninklijk Nederlands Instituut voor Onderzoek der Zee (NIOZ), 
Texel, the Netherlands}
\newcommand{\nNIO}{40}
\newcommand{\tNOA}{NOA/NESTOR, Pylos, Greece}
\newcommand{\nNOA}{41}
\newcommand{\tUSh}{University of Sheffield, United Kingdom}
\newcommand{\nUSh}{42}
\newcommand{\tTec}{Tecnomare, Ven, Italy}
\newcommand{\nTec}{43}
\newcommand{\tUUt}{University of Utrecht, the Netherlands}
\newcommand{\nUUt}{44}
\newcommand{\tPUV}{Institut d'Investigaci\'o per a la Gesti\'o integrada de les 
Zones Costaneres, Universitat Polit\`ecnica Val\`encia, Gandia, Spain}
\newcommand{\nPUV}{45}

\author{%
\begin{center}
{\bf The KM3NeT Collaboration}\\[3.mm]
 S. Adri\'an-Mart\'inez$^{\nPUV}$,
 M.~Ageron$^{\nMar}$,
 J.A.~Aguilar$^{\nIFI}$,
 F.~Aharonian$^{\nDub}$,
 S.~Aiello$^{\nICt}$,
 A.~Albert$^{\nUHA}$,
 M.~Alexandri$^{\nHCM}$,
 F.~Ameli$^{\nIRo}$,
 E.G.~Anassontzis$^{\nUAt}$,
 M.~Anghinolfi$^{\nIGe}$
% ...
\end{center}
{\small
\setlongtables
\begin{longtable}{rp{12.3cm}}
$^{\nUAb}$&\tUAb\\
$^{\nUAm}$&\tUAm\\
$^{\nAPC}$&\tAPC\\
$^{\nUAt}$&\tUAt\\
$^{\nBar}$&\tBar\\
$^{\nCIR}$&\tCIR\\
$^{\nCLS}$&\tCLS\\
$^{\nCNR}$&\tCNR\\
$^{\nMar}$&\tMar\\
$^{\nCyp}$&\tCyp\\
$^{\nDub}$&\tDub\\
$^{\nUEr}$&\tUEr\\
$^{\nUHA}$&\tUHA\\
$^{\nHCM}$&\tHCM\\
$^{\nHOU}$&\tHOU\\
$^{\nIFI}$&\tIFI\\
$^{\nIFR}$&\tIFR\\
$^{\nIBa}$&\tIBa\\
$^{\nIBo}$&\tIBo\\
$^{\nICa}$&\tICa\\
$^{\nICt}$&\tICt\\
$^{\nUCt}$&\tUCt\\
$^{\nIGe}$&\tIGe\\
$^{\nLNS}$&\tLNS\\
$^{\nLNF}$&\tLNF\\
$^{\nINa}$&\tINa\\
$^{\nUNa}$&\tUNa\\
$^{\nIPi}$&\tIPi\\
$^{\nIRo}$&\tIRo\\
$^{\nURo}$&\tURo\\
$^{\nING}$&\tING\\
$^{\nUSt}$&\tUSt\\
$^{\nISS}$&\tISS\\
$^{\nUKi}$&\tUKi\\
$^{\nKVI}$&\tKVI\\
$^{\nULe}$&\tULe\\
$^{\nULi}$&\tULi\\
$^{\nNCS}$&\tNCS\\
$^{\nNik}$&\tNik\\
$^{\nNIO}$&\tNIO\\
$^{\nNOA}$&\tNOA\\
$^{\nUSh}$&\tUSh\\
$^{\nTec}$&\tTec\\
$^{\nUUt}$&\tUUt\\
$^{\nPUV}$&\tPUV\\
$^a$&also at Universit\'e Paris-Sud, 91405 Orsay Cedex, France\\
$^b$&now at Universit\"at W\"urzburg, Germany\\
$^c$&on leave of absence at Humboldt University, Berlin, Germany\\
$^d$&now at Dept.\ of Physics and Wisconsin IceCube Particle Astrophysics Center, 
University of Wisconsin, Madison, WI 53706, USA\\
$^e$&also at Moscow State University, Skobeltsyn Institute of Nuclear Physics, Moscow, Russia\\
$^f$&now at Bogolyubov Institute for Theoretical Physics, Ukraine\\
$^g$&now at Laboratoire d'Annecy-le-Vieux de physique des particules (LAPP), France\\
$^h$&now at Institut for Cosmic Ray Research, University of Tokyo, Japan\\
$^i$&now at Universit\'e Paris VI, Laboratoire de Physique Nucl\'eaire et de Hautes Energies (LPNHE), France\\[2.mm]
$^\star$&{\normalsize corresponding author, e-mail address: coniglione@lns.infn.it}
\end{longtable}
}
}

\begin{abstract}
A recent analysis of the Fermi Large Area Telescope data provided evidence for
a high-intensity emission of high-energy gamma rays with a $E^{-2}$ spectrum
from two large areas, spanning $50^\circ$ above and below the Galactic centre
(the ``Fermi bubbles''). A hadronic mechanism was proposed for this gamma-ray
emission making the Fermi bubbles promising source candidates of high-energy
neutrino emission. In this work Monte Carlo simulations regarding the
detectability of high-energy neutrinos from the Fermi bubbles with the future
multi-km$^3$ neutrino telescope KM3NeT in the Mediterranean Sea are presented. 
Under the hypothesis that the gamma-ray emission is completely due to hadronic
processes, the results indicate that neutrinos from the bubbles could be
discovered in about one year of operation, for a neutrino spectrum with a cutoff
at $100\,$TeV and a detector with about $6\,\mathrm{km}^3$ of instrumented
volume. The effect of a possible lower cutoff is also considered.
\end{abstract}

\end{frontmatter}


\section{Introduction}

In the last decade a new generation of telescopes revealed a large variety of
astrophysical high-energy gamma-ray sources. Telescopes such as Fermi
\cite{FERMI}, HESS \cite{HESS}, VERITAS \cite{VERITAS} and MAGIC \cite{MAGIC}
can identify sources of gamma rays with energies from about $20\,$MeV to about 
$100\,$TeV with sufficient angular resolution to study the source morphology of
extended Galactic sources. They can also measure the gamma-ray energy spectrum
with high precision. From these measurements the presence of cosmic acceleration
processes has been confirmed in a large variety of known and unknown galactic
and extragalactic sources \cite{AhaReport,MilagroObs}.

For the full understanding of the underlying mechanisms other probes that can
complement the information from gamma-ray detection are needed. As neutrinos are
mainly produced via proton-proton ($pp$) and proton-gamma ($p\gamma$)
interactions, they are an unambiguous signature of hadronic acceleration.

Neutrino telescopes have been constructed, and more sensitive ones are proposed,
to detect these high-energy neutrinos. Neutrino flux estimates indicate that
detectors of km$^{3}$-scale instrumented volume are required \cite{Kappes}. The
IceCube \cite{IceCube} detector located at the South Pole, about
$1\,\mathrm{km}^3$ in size, is taking data in its final configuration since
2011. In the northern hemisphere ANTARES is the largest operating detector
located in the Mediterranean Sea $40\,$km off the French coast near Toulon
\cite{ANTARES}. It has an instrumented volume of about $0.01\,\mathrm{km}^3$ and
it is taking data in its final configuration since 2008. The KM3NeT consortium
\cite{KM3NeT} proposes the construction of a multi-km$^3$ telescope in the
Mediterranean Sea (KM3NeT). From their respective locations, KM3NeT and ANTARES
will detect upward-going neutrinos from about $3.5\pi\,$sr of the sky, including
the Galactic centre and most of the Galactic plane, where many TeV gamma-ray
sources are located \cite{AhaReport,Kappes}.

In this work the response of the KM3NeT telescope to neutrinos from the Fermi
bubbles (see below) is studied with Monte Carlo simulations under the hypothesis
that a hadronic mechanism is responsible for the gamma-ray emission. In
particular, the neutrino flux that could be discovered by KM3NeT is determined
as a function of the number of observation years.

\section{The Fermi bubbles}

A recent analysis of Fermi-LAT data \cite{FermiBubbles} revealed an intense
gamma-ray emission from two large areas above and below the Galactic centre. The
detected gamma-ray emission has the following characteristics:
\begin{itemize}
\item 
The emission areas are symmetric with respect to the Galactic plane and extend
up to 50\,degrees (10\,kpc) north and south the Galactic centre, with a width
of 40\,degrees in Galactic longitude.
\item 
The gamma energy spectrum, measured from $\sim1\,$GeV to $\sim100\,$GeV, is
compatible with a power-law spectrum described by $E^{2}\,d\Phi_{\gamma}/dE\approx3\mathrm{-}6
\times10^{-7}  \,\mathrm{GeV\,\mathrm{cm}^{-2}\,
\mathrm{s}^{-1}\,\mathrm{sr}^{-1}}$.
\item 
The emission is homogeneous within the bubbles. No significant differences were
found between the northern and the southern bubble.
\item 
The edges of the bubbles seem to be correlated with ROSAT X-ray maps at
$1.5\mathrm{-}2\,$keV, while the inner parts are correlated with the
hard-spectrum microwave excess known as WMAP haze \cite{FermiBubbles,Dobler
2010,Dobler 2011}.
\end{itemize}

Several mechanisms have been suggested in the literature to explain this
gamma-ray emission. Most of these explanations rely on leptonic processes and
include inverse Compton scattering by electrons, either produced by long-lasting
energy injections near the Galactic centre \cite{FermiBubbles} or accelerated
through the second-order Fermi process generated by turbulent plasma throughout
the entire bubble \cite{Mertsch}. Other processes such as recent transient AGN
activity near the Galactic centre \cite{Guo}, millisecond pulsars
\cite{Malyshev} and dark matter annihilation \cite{Dobler 2011DM} have also been
put forward.

Currently the observed features cannot be fully explained by these
predominantly leptonic processes, which has led to a proposal for an underlying hadronic 
process \cite{Crocker}. A cosmic ray population associated with long-time-scale
star formation in the Galactic centre (of the order of $10^{10}\,$years) was hypothesised to have been
injected into the bubbles where it interacts with the ambient matter and
produces high-energy gamma rays through $\pi^0$ decay. Pair production is
responsible for the generation of secondary electrons producing the synchrotron
radiation observed as microwaves. In this scenario neutrinos are produced
through the decay of charged pions. The leptonic and hadronic mechanisms require
very different galactic diffusion characteristics.

Recently the morphological and spectral characteristics of the WMAP haze have
been reviewed \cite{Dobler 2011}, with important implications for the possible
production scenarios. The author considers all the proposed models to explain
the gamma and the WMAP haze emission and concludes that hybrid emission scenarios
will likely be required.
 
The detection of high-energy neutrinos can distinguish between hadronic and
leptonic models. Only if a hadronic process is totally or partially responsible
for the production of gamma rays from the Fermi bubbles, neutrinos will also be
produced. The KM3NeT telescope, to be located in the deep Mediterranean Sea, will be
the ideal instrument for the observation of neutrinos from the Fermi bubbles.

\section{The detector}

The KM3NeT consortium aims at installing a deep-sea research infrastructure
hosting a multi-cubic-kilometre high-energy neutrino detector in the
Mediterranean Sea. During the Design Study\footnote{Supported by the EU in FP6,
contract n$^\circ$ 011937}, several technical solutions were investigated and
the results reported in the Technical Design Report \cite{TDR}. During a
Preparatory Phase\footnote{Supported by the EU in FP7, grant agreement n$^\circ$
212525}, a final design concept was defined. Currently prototyping activities
are in progress to prepare for the construction phase.

The detection principle relies on the detection of Cherenkov light induced by
secondary charged particles produced in neutrino-nucleus interactions inside or
near the detector. The Cherenkov light is detected by photomultiplier tubes
(PMTs) contained in glass spheres that are designed to resist the hydrostatic
pressure of the deep-sea environment. These instrumented spheres are called
Optical Modules (OMs).

The geometry of the detector \cite{KooVLVNT11} simulated in this work consists
of a three-dimensional array of OMs attached to vertical structures (Detection
Units, DUs). An array of DUs constitutes a detector building block. Several
building blocks form the full detector of about 300~DUs. A DU, which is anchored
to the sea floor and kept upright by a submerged buoy, consists of horizontal
bars equipped with 2~OMs, one at each end. Adjacent bars are oriented
orthogonally to each other. The DUs are connected to shore by an electro-optical
cable. Each OM consists of 31 3-inch PMTs \cite{MultiPMT} housed inside a
17-inch pressure-resistant glass sphere covering almost a $4\pi$ field of view.

For this work, a detector of 308~DUs (two building blocks of 154~DUs each,
arranged uniformly in a circular area) with an average separation between
adjacent DUs of $180\,$m has been simulated. Each simulated DU consists of
20~bars of $6\,$m length with a vertical spacing of $40\,$m (40~OMs per DU). The
total instrumented detector volume is about $6\,\mathrm{km}^3$.

\section{Monte Carlo simulation}

In order to study the sensitivity of the KM3NeT telescope for the detection of
neutrinos from the Fermi bubbles, we use a Monte Carlo simulation framework. The
framework is based on the ANTARES software \cite{ANTAREScode}, modified for a
$\mathrm{km}^3$-scale detector and using OM properties appropriate for KM3NeT. 
The simulation chain consists of the generation of muon neutrinos from the
bubbles, the generation of atmospheric muon and neutrino backgrounds, the
neutrino charged-current interactions, the propagation of the produced muons in
rock and sea water, the generation of Cherenkov light, the $^{40}\mathrm{K}$
background and the digitisation of the PMT signals. Optical properties of the
sea water and the PMT characteristics are taken into account in the simulation. 
The depth and the optical water properties measured at the Sicilian Capo Passero
site have been used \cite{TDR}. Background light due to the presence of
$^{40}\mathrm{K}$ in salt water has been simulated adding an uncorrelated hit
rate of $5\,$kHz per PMT and a time-correlated hit rate of $500\,$Hz per OM (two
coincident hits in different PMTs inside the same OM). These parameters have
been estimated with a complete simulation based on GEANT4 \cite{GEANT4}. The
$5\,$kHz of uncorrelated hit rate is consistent with the baseline of $50\,$kHz
measured with the 10-inch PMTs of the ANTARES experiment
\cite{ANTARESfirstline}.

The muon track direction is reconstructed from the simulated arrival times of
Cherenkov photons and the PMT positions. The reconstruction algorithm is based
on a hit selection and on a maximum likelihood fit that uses probability density
functions for the photon arrival times at the PMTs. In addition to the positions
and track directions, the number of hits used for the reconstruction
($N_\mathrm{hit}$) and a track fit quality parameter ($\Lambda$) are given as
output. The $\Lambda$ parameter is determined from the likelihood and from the
number of compatible track solutions found by the algorithm and is used to
reject badly reconstructed events \cite{Heijboer}. The $N_\mathrm{hit}$
parameter is correlated with the muon energy.

\subsection{Neutrinos from the Fermi bubbles}

Muon neutrinos from the Fermi bubbles were generated homogeneously within
two circular regions of $19^\circ$ radius around two positions in the sky at
the equatorial coordinates declination $\delta=-15^{\circ}$ and right ascension
$\alpha=243^{\circ}$ for the northern bubble and $\delta=-44^{\circ}$ and
$\alpha=298^{\circ}$ for the southern bubble. The simulated neutrino energy is
between $10^{2}$ and $10^{8}\,$GeV.

\begin{figure}[!htb]
\centering
\includegraphics[scale=0.7]{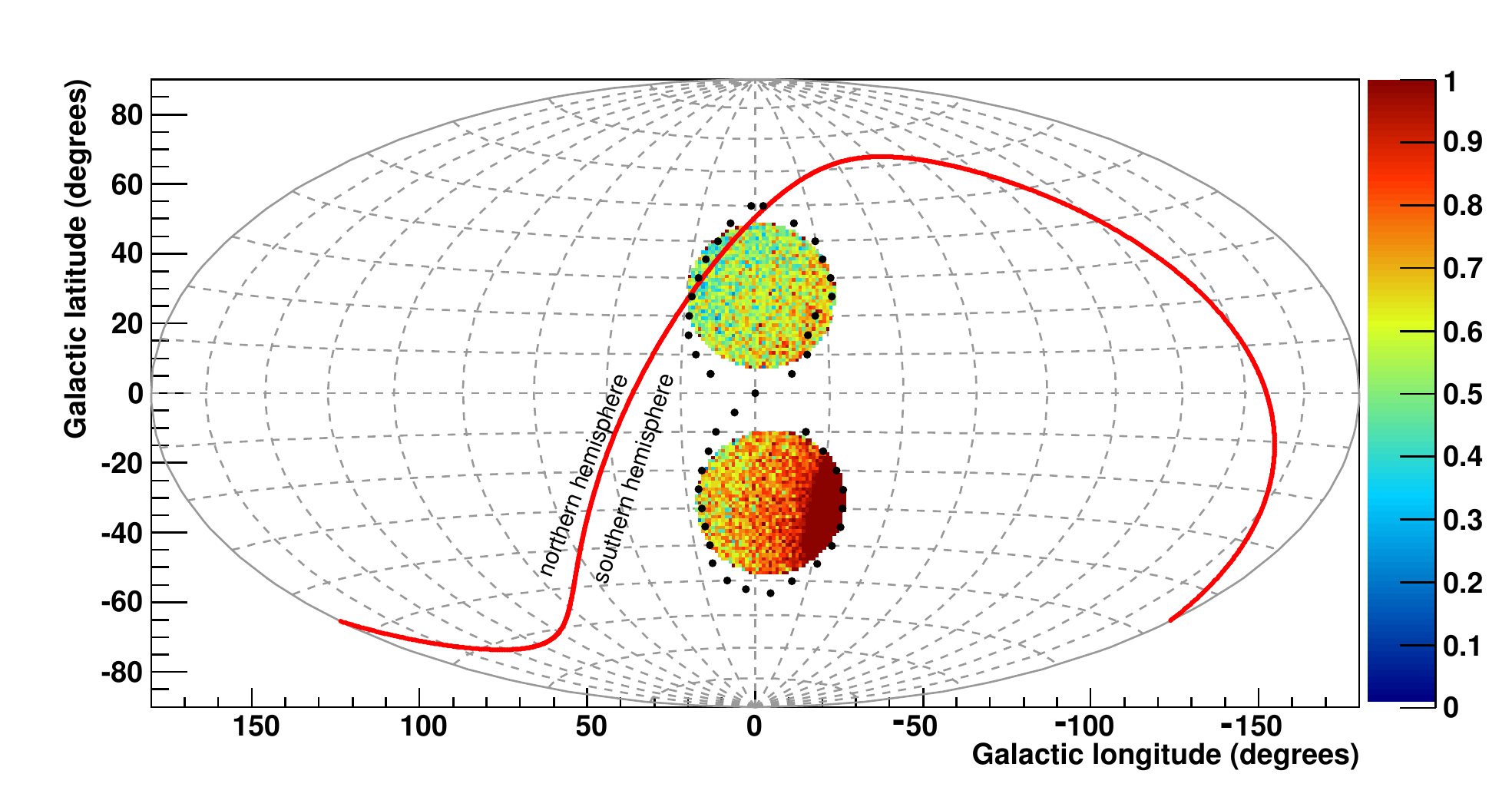}
\caption{
 Galactic coordinates of simulated neutrino from the Fermi bubbles. 
The color code indicates the fraction of Monte Carlo neutrinos that
are up-going in local detector coordinates. The measured bubble edges
(\cite{FermiBubbles} Table 1) are also reported (black points). The red line
represents the separation line between the northern and southern hemisphere.}
\label{fig:GenEvents}
\end{figure}

Figure \ref{fig:GenEvents} shows the percentage of Monte Carlo neutrinos that
are up-going in local detector coordinates. Since only up-going events can be
unambiguously classified as neutrino candidates, Fig.~\ref{fig:GenEvents}
represents the Fermi bubbles' visibility for a neutrino telescope in the
Mediterranean Sea. Black points are the edges of the two bubbles as reported in
\cite{FermiBubbles}. The average visibility for a detector located at a latitude
of $36^{\circ}\;16'\,$N is $58\%$ of the time for the northern bubble and $80\%$
for the southern bubble. We note that for IceCube only a small fraction of the
Fermi bubbles lies below the horizon (solid red line in
Fig.~\ref{fig:GenEvents}), leading to a significantly reduced sensitivity
compared to a similar detector in the Mediterranean Sea \cite{IceFermiBubble}.

The generated Monte Carlo events can be weighted to reproduce different assumed
neutrino spectra. The general energy dependence of the neutrino flux used in
this analysis is a power law spectrum with a spectral index of $-2$ and an
exponential cutoff:
\begin{equation}
 \frac {d\Phi_\nu^{E_c}}{dE} = K_{0\nu} \cdot E^{-2} \cdot e^{-E/E_c}\,,
 \label{eq:E2}
\end{equation}
where $E_c$ is the cutoff energy. For this analysis we consider three cases:
$d\Phi^{\infty}_{\nu}/dE$, i.e.\ a pure power law spectrum with spectral index
$-2$, as well as $d\Phi^{100}_{\nu}/dE$ and $d\Phi^{30}_{\nu}/dE$ with cutoffs
at $100\,$TeV and $30\,$TeV, respectively. The cutoff at $E_c=100\,$TeV is
consistent with an assumed cutoff of the proton energy distribution in the Fermi
bubbles at a few PeV (corresponding to the position of the ``knee'' in the
cosmic ray energy spectrum). The second cutoff value considered, $E_c=30\,$TeV,
is more conservative and does not require the assumption that the Fermi bubbles
populate the cosmic ray spectrum up to the knee.

Under the hypothesis that the source is transparent to gamma rays and that the
mechanism responsible for the gamma-ray emission is hadronic, the neutrino
spectrum was estimated from the measured gamma-ray spectrum following the
prescription described in \cite{Vissani}. The measured Fermi bubble gamma-ray
spectrum is consistent with a $E^{-2}$ shape and a normalisation
$K_{0\gamma}\approx4\times 10^{-7}\;\mathrm{GeV\,cm^{-2}\,s^{-1}\,sr^{-1}}$. 
Using this gamma-ray flux and taking into account the $0.7\,$sr solid angle of
the two simulated bubbles, we estimated the corresponding muon neutrino plus 
antineutrino flux to have $K_{0\nu}\approx1\times10^{-7}\;\mathrm{GeV\,cm^{-2}\,s^{-1}}$.

\subsection{Atmospheric muons}

Cosmic rays entering the atmosphere produce extensive air showers that contain
high-energy muons. Although the sea water above the detector serves as a shield,
many of these muons reach the detector. Therefore down-going tracks are excluded
from the analysis. However, atmospheric muons misreconstructed as up-going
remain a significant background. Simulating atmospheric muon background in a
large detector requires huge amounts of CPU time and data storage. Atmospheric
muons were generated with the fast MUPAGE code \cite{MUPAGE}. This code, which
is based on a full Monte Carlo simulation of primary cosmic ray interactions and
shower propagation in the atmosphere, provides a parameterisation of the
underwater flux of atmospheric muons including also multi-muon events (``muon
bundles'').

Atmospheric muons were generated in the range $1\,\mathrm{TeV}\leq
E_{b}<10\,$TeV, where $E_{b}$ is the sum of the energies of all single muons in
the bundle. This sample is statistically equivalent to a live time of 2\,hours. 
To efficiently increase statistics in the high-energy region an additional
sample equivalent to 8~days of live time was generated with $E_{b}\geq10\,$TeV. 
The Monte Carlo samples were then reweighted to the relevant experimental live
time.

\subsection{Atmospheric neutrinos}

A large number of secondary particles is produced in a cosmic ray interaction in
the atmosphere. The production of pions and kaons and their subsequent decay
chains produce a large flux of atmospheric neutrinos, called ``conventional
flux'', with an energy distribution proportional to $E_{\nu}^{-3.7}$ at neutrino
energies $E_{\nu}$ larger than about $1\,$TeV. The production of neutrinos from
heavy quark is responsible for a high-energy component ($E_{\nu}\geq10\,$TeV) in
the atmospheric neutrino spectrum. This contribution, called ``prompt'', is not
well known, but several predictions are available \cite{Costa,Enberg}. The
atmospheric neutrino flux is an irreducible background for the detection of
neutrinos of cosmic origin.

The atmospheric muon neutrino and antineutrino background was generated in the
energy range $10^{2}\,\mathrm{GeV}\leq E_{\nu}\leq 10^{8}\,\mathrm{GeV}$
and over the full solid angle. The events are weighted to reproduce the conventional
atmospheric neutrino flux following the Bartol model \cite{Bartol}. The
uncertainty on the Bartol normalisation factor is taken to be about 25--30\%
\cite{ANTdiffuse}. A prompt contribution is also taken into account through the
event weights. The models in \cite{Costa,Enberg} have been considered and the
model with the highest neutrino flux, corresponding to the highest prediction of
the Recombination Quark Parton Model (RQPM) \cite{Bugaev}, has been used in the
present analysis.

\section{Flux discovery potential}

The present analysis was restricted to events reconstructed as up-going and
located within $19^\circ$ around the centre of each Fermi bubble. The bubbles
are considered as discovered if the number of detected events in a given
detector live time has a probability of $\alpha = 2.85\times10^{-7}$ or less to
originate purely from background in $1-\beta = 50\%$ of all experiments. This
corresponds to a significance of $5\sigma$ (area of the one-sided Gaussian
tail).

The signal flux required to claim a discovery is calculated from the simulated
average number of background events, $\langle n_\mathrm{back}\rangle$. First,
the minimum (critical) number of events, $n_0$, that satisfies
\begin{equation}
\sum_{n_\mathrm{obs}=n_0}^{\infty} P(n_\mathrm{obs}| \langle n_\mathrm{back} \rangle) < \alpha
\end{equation}
is determined, where $P (n_\mathrm{obs} | \langle n_\mathrm{back}\rangle)$ is
the Poisson probability for observing $n_\mathrm{obs}$ events given $\langle
n_\mathrm{back}\rangle$. The value $n_0$ is the minimum number of events
required to claim a deviation from the background-only hypothesis with a
statistical significance defined by the $p$-value $\alpha$. The confidence level
C.L.\ is related to the $p$-value via $\mathrm{C.L.}=1-\alpha$.

In case of the presence of a signal of strength $n_\alpha$, the probability to
observe $n_0$ or more events is related to the statistical power $(1-\beta)$ by
\begin{equation}
\sum_{n_\mathrm{obs}=n_0}^\infty P(n_\mathrm{obs}| \langle n_\alpha \rangle + 
\langle n_\mathrm{back}\rangle) = 1 - \beta\,.
\end{equation}
The signal strength $n_\alpha(\langle n_\mathrm{back}\rangle)$ resulting from (3) would lead to an observation with a $p$-value less
than $\alpha$ in a fraction $(1-\beta)$ of the experiments.

In order to determine the minimum flux $\phi_\alpha$ needed for a discovery (discovery
flux), the cuts on $\Lambda$ and $N_\mathrm{hit}$ were varied and the minimum in
the Model Discovery Potential (MDP) \cite{Aharens}, defined as $\mathrm{MDP} =
n_\alpha(\langle n_\mathrm{back}\rangle) / \langle n_s \rangle$ was sought. 
Here, $\langle n_s \rangle$ is the number of signal events resulting from the
model flux (\ref{eq:E2}) with $K_{0\nu} \approx 1 \times 10^{-7}\,
\mathrm{GeV\,cm^{-2}\,s^{-1}}$ after cuts. The discovery flux is then related
to the MDP by
\begin{equation}
\phi_\alpha = K_{0\nu} \cdot \mathrm{MDP} = 
K_{0\nu} \cdot \frac{n_\alpha(\langle n_\mathrm{back}\rangle)}{\langle n_s \rangle}\,.
\end{equation}

\begin{figure*}[!htb]
\centering
\includegraphics[scale=0.8]{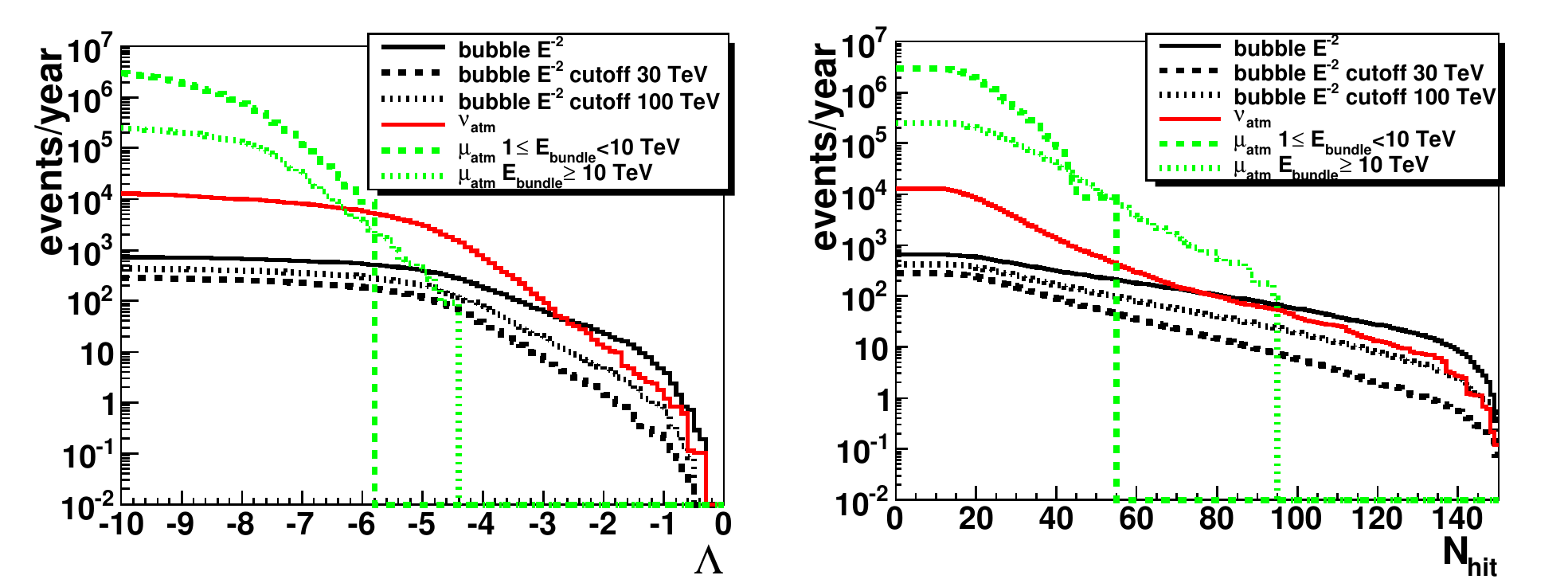}
\caption{
Cumulative distribution of the number of events per year reconstructed as up-going
in the bubble region as a function of $\Lambda$ (left panel) and of
$N_\mathrm{hit}$ (right panel) for the different Monte Carlo
samples. For the right panel a cut $\Lambda \geq -10$ was applied.}
\label{fig:LambdaNhit}
\end{figure*}

Figure~\ref{fig:LambdaNhit} shows the cumulative distributions (as functions of $\Lambda$ and $N_\mathrm{hit}$) of the numbers of
events per year, reconstructed as up-going in the bubble region, for the
different Monte Carlo samples. 
Figure~\ref{fig:LambdaNhit} demonstrates that most of the events due to
atmospheric muons that are misreconstructed as up-going can be rejected by imposing
an appropriate cut on $\Lambda$, while an appropriate cut on $N_\mathrm{hit}$
helps in the rejection of atmospheric neutrino background events
\cite{ANTpoint}.

\begin{figure*}[!htb]
\centering
\includegraphics[scale=0.8]{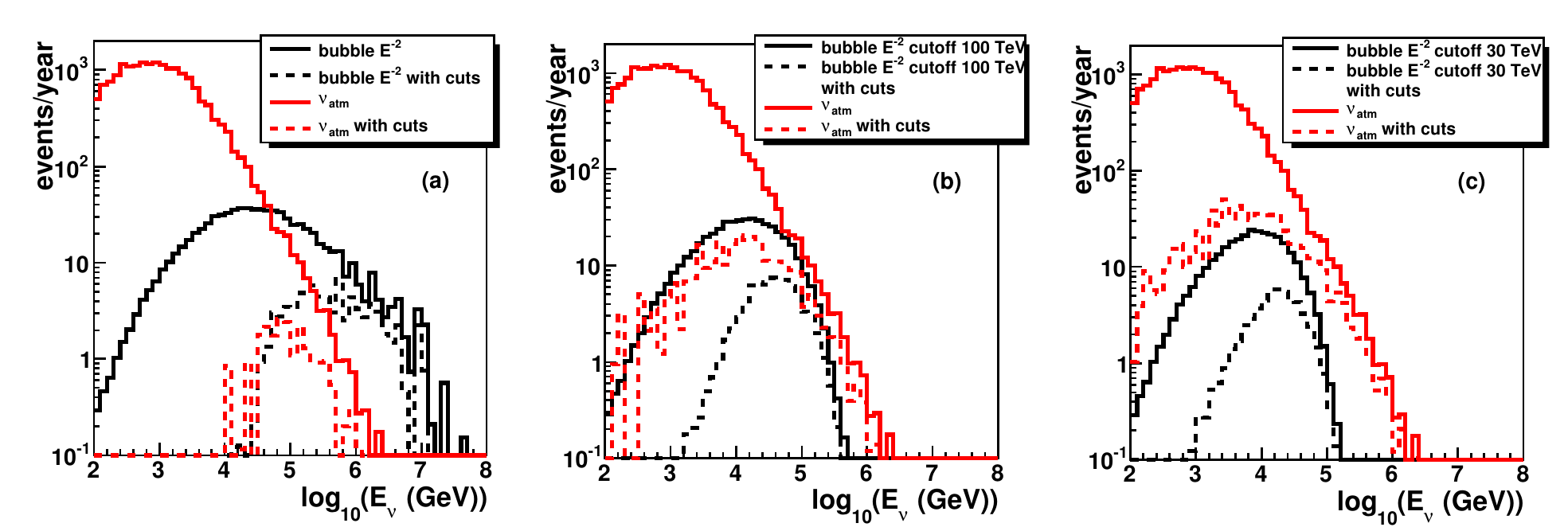}
\caption{
Number of events per year reconstructed as up-going in the bubble region as a
function of the simulated neutrino energy, for the signal and the atmospheric
neutrino background with and without the cuts required to minimise the MDP
($5\sigma$ C.L., $50\%$ probability). The assumed normalisation factor is
$K_{0\nu} = 1 \times 10^{-7}\,\mathrm{GeV}\,\mathrm{cm}^{-2}\,
\mathrm{s}^{-1}$. The bin width in $\log_{10}(E_{\nu})$ is 0.1.}
\label{fig:EnergySpectra}
\end{figure*}

The cuts required to minimise the MDP vary for the different flux assumption and
are given in Tab.~\ref{tab:Events}. Figure~\ref{fig:EnergySpectra} shows the
spectrum of reconstructed signal and background up-going events per year in the
region of the bubbles as a function of the simulated neutrino energy. The signal
energy spectrum from the bubbles, after the cuts, is centred at about $500\,$TeV
for $d\Phi^{\infty}_{\nu}/dE$ (see Eq.~(\ref{eq:E2})) (a), at about $40\,$TeV
for $d\Phi^{100}_{\nu}/dE$ (b) and at about $25\,$TeV for $d\Phi^{30}_{\nu}/dE$
(c). The numbers of events at the reconstruction level and after the cuts in one
year of data taking are reported in Tab.~\ref{tab:Events} for a source with a
flux normalisation factor of $K_{0\nu} = 1 \times 10^{-7}\,\mathrm{GeV}\,
\mathrm{cm}^{-2}\,\mathrm{s}^{-1}$.

\setcounter{table}{0}

\begin{table}[!htb]
\begin{center}
\begin{tabular}{|c|c|c|c|c|c|c|} 
\hline
Spectrum & & \textbf{$\nu_\mathrm{sig}$} & $\nu_\mathrm{atm}$ &$\mu_\mathrm{atm}$ & $\Lambda$ &
$N_\mathrm{hit}$\\
\hline
\multirow{2}*{$\propto  E^{-2}$} & reco level & 770 & 17707 & $8\times10^{6}$& &\\
\cline{2-7}
& cut level &  80 &  21 & 0 & $-5$ &129\\
\hline
\multirow{2}*{$\propto  E^{-2} e^{-E/100 \,\mathrm{TeV}}$} & reco level & 474 & 17707 & $8\times10^{6}$ & &\\
\cline{2-7}
& cut level & 77 & 282 & 0 & $-4.7$ & 56\\
\hline
\multirow{2}*{$\propto  E^{-2} e^{-E/30 \,\mathrm{TeV}}$} & reco level & 315 & 17707 & $8\times10^{6}$ & &\\
\cline{2-7}
& cut level &  53 & 658 & 0 & $-4.8$ & 43\\
\hline
\end{tabular}
\end{center}
\caption{
Expected numbers of events in one year of data taking for the three samples of
simulated events: neutrinos from the Fermi bubbles ($\nu_\mathrm{sig}$),
neutrinos from atmospheric background (conventional plus prompt RQPM)
($\nu_\mathrm{atm}$) and muons from atmospheric background ($\mu_\mathrm{atm}$). 
The cut values ($\Lambda$ and $N_\mathrm{hit}$) and the numbers of events at the
reconstruction level (reco level) and after the cuts (cut level) are also
given. The cuts are those optimised for a discovery at $5\sigma$ C.L., $50\%$
probability, for a source with a normalisation factor of $K_{0\nu} = 1 \times
10^{-7} \mathrm{GeV\, cm^{-2}\,s^{-1}}$.}
\label{tab:Events}
\end{table} 

With the small simulated live time of the atmospheric muon samples the number of
events per year at cut level due to the atmospheric muons is zero (see
Tab.~\ref{tab:Events}). By extrapolating in these data sets the trends of the
numbers of events per year as a function of $\Lambda$ and $N_\mathrm{hit}$
(Fig.~\ref{fig:LambdaNhit}), it was estimated that, in the cases of source neutrino
spectra with cutoffs, the contribution of atmospheric muons to the background
does not exceed $25\%$ of the atmospheric neutrino background events reported in
Tab.~\ref{tab:Events}. A $25\%$ increase in the number of background events
corresponds to an increase of $10\%$ in the flux limits. In the case of a pure
power-law spectrum the contribution of atmospheric muons to the background is
negligible.

\begin{figure}[!htb]
\centering
\includegraphics[scale=.6]{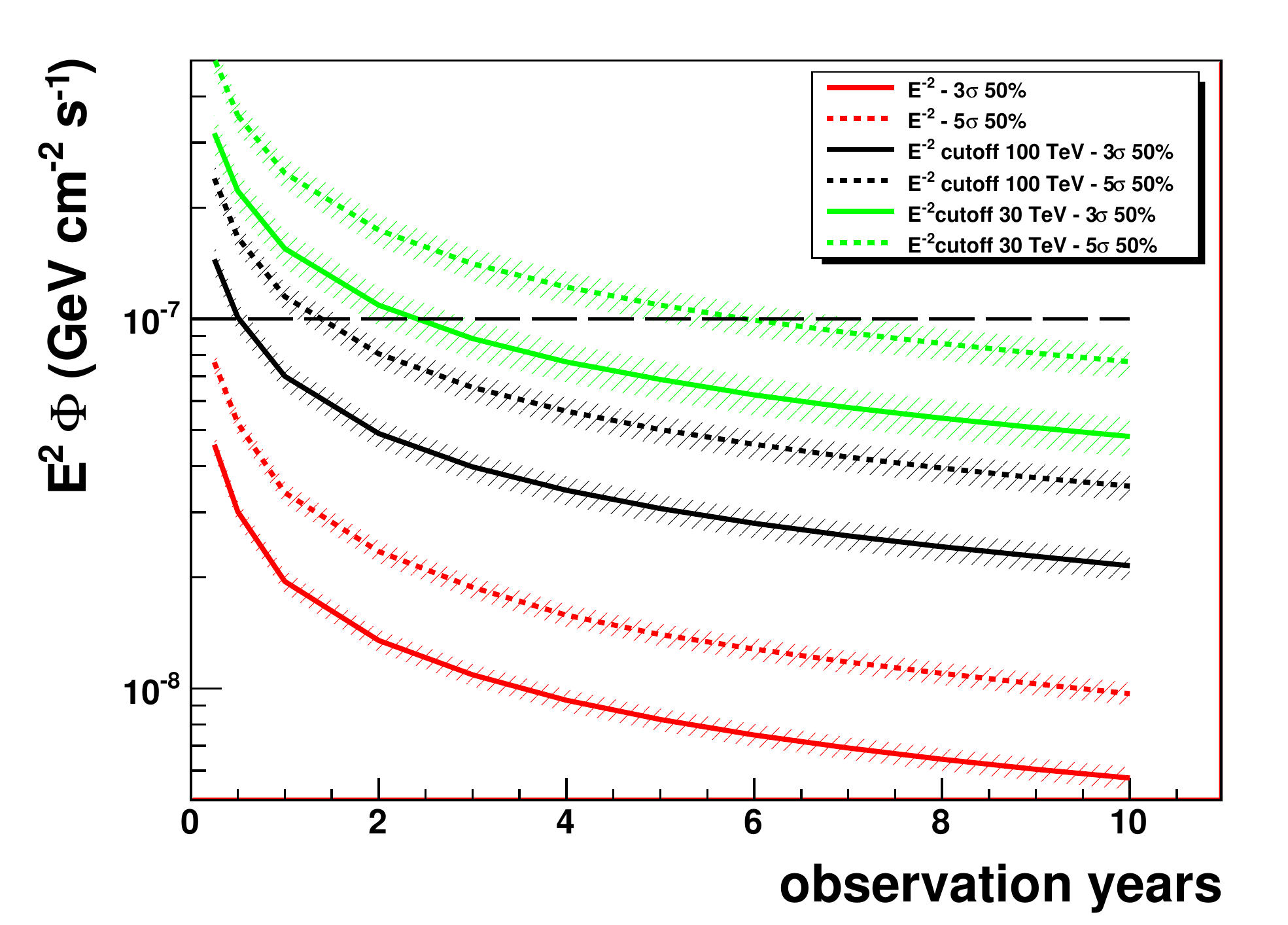}
\caption{
Discovery fluxes as functions of the observation years for $5\sigma$ C.L.,
$50\%$ probability and $3\sigma$ C.L., $50\%$ probability, for the three
neutrino spectra assumed. The bands represent the variation due to the
uncertainty on the normalisation factor of the conventional Bartol neutrino
flux. The long-dashed line indicates the predicted neutrino flux estimated in Sec.~4.1.}
\label{fig:DiscFlux}
\end{figure}

In Fig.~\ref{fig:DiscFlux} the discovery fluxes at $5\sigma$ C.L., $50\%$
probability and $3\sigma$ C.L., $50\%$ probability for the neutrino spectra
considered (see Eq.~(\ref{eq:E2})) are shown as functions of the observation
time. The variation in the discovery fluxes due to the uncertainties in the
atmospheric neutrino flux have been evaluated taking into account only the
uncertainty on the normalisation factor of the conventional Bartol flux. The
RQPM prompt model component contributes with about $46\%$ and $10\%$ in the
number of neutrino background events after the cuts quoted in
Tab.~\ref{tab:Events} for the pure $E^{-2}$ spectrum and for the exponential
cutoff spectra, respectively. A $25\%$ uncertainty in the conventional Bartol
atmospheric neutrino flux corresponds to a variation in the discovery flux of
about $7\%$ for $d\Phi^{\infty}_{\nu}/dE$ (see Eq.~(\ref{eq:E2})) and $10\%$ for
$d\Phi^{100}_{\nu}/dE$ and $d\Phi^{30}_{\nu}/dE$. These variations are shown in
Fig.~\ref{fig:DiscFlux} as bands around the discovery-flux curves.

If the neutrino normalisation factor is of the order of $1\times 10^{-7}\,
\mathrm{GeV\,cm^{-2}\,s^{-1}}$ the discovery of neutrinos from the Fermi
bubbles is expected in about one year of data taking for a $E^{-2}$ neutrino
spectrum with a cutoff at $100\,$TeV. The first evidence ($3\sigma$ C.L., $50\%$
probability) could be obtained after a few months of data taking. For the more
severe cutoff at $30\,$TeV the discovery is predicted to be achieved in about
5.5\,years and first evidence in about 2.5\,years.

\begin{figure*}[!htb]
\centering
\includegraphics[scale=0.4]{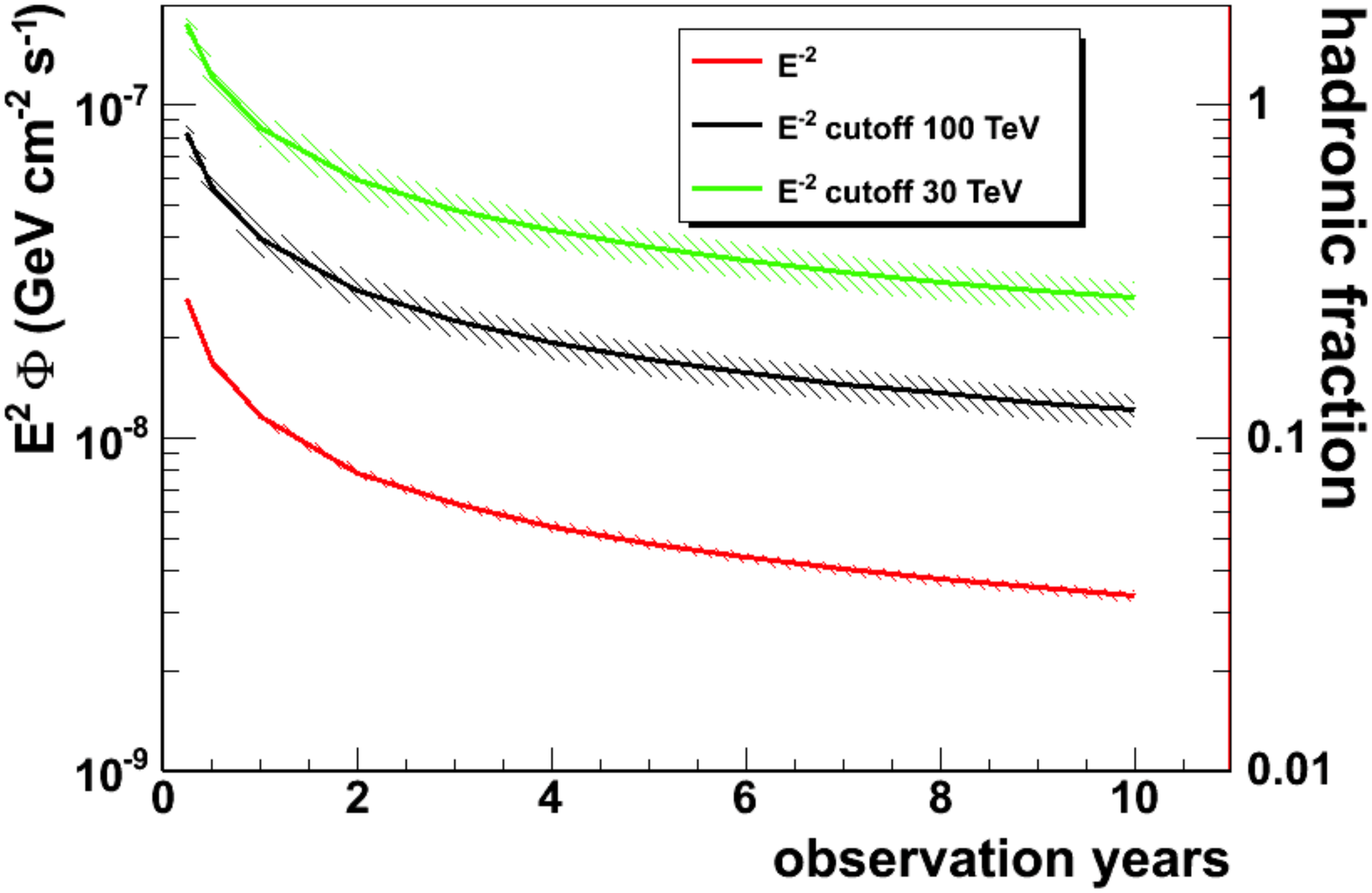}
\caption{
Sensitivity at $90\%$ C.L. as a function of the observation years for the three neutrino spectra assumed.
The right vertical scale indicates the upper limit on the fraction of hadronic emission (see text).
The bands represent the variation due to the uncertainty on the normalisation factor of the conventional Bartol neutrino
flux.}
\label{fig:SensFlux_km3net}
\end{figure*}

\begin{figure*}[!htb]
\centering
\includegraphics[scale=0.6]{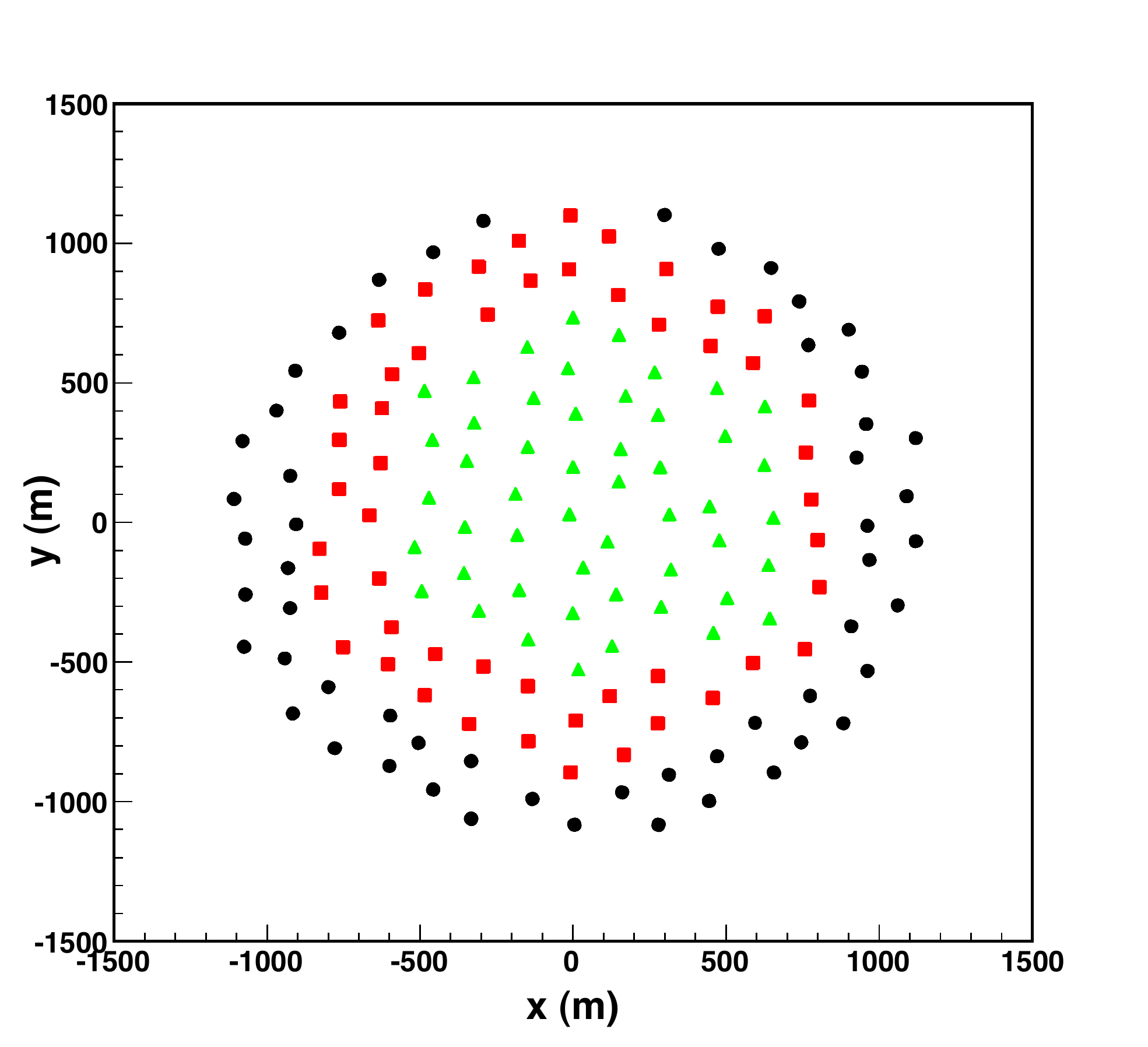}
\caption{DU positions for a single block of detectors composed of 154 DUs (black circles, red squares and green triangles), 100 DUs (red squares and green triangles) and 50 DUs (green triangles).}
\label{fig:Footprint}
\end{figure*}

\begin{figure*}[!htb]
\centering
\includegraphics[scale=0.8]{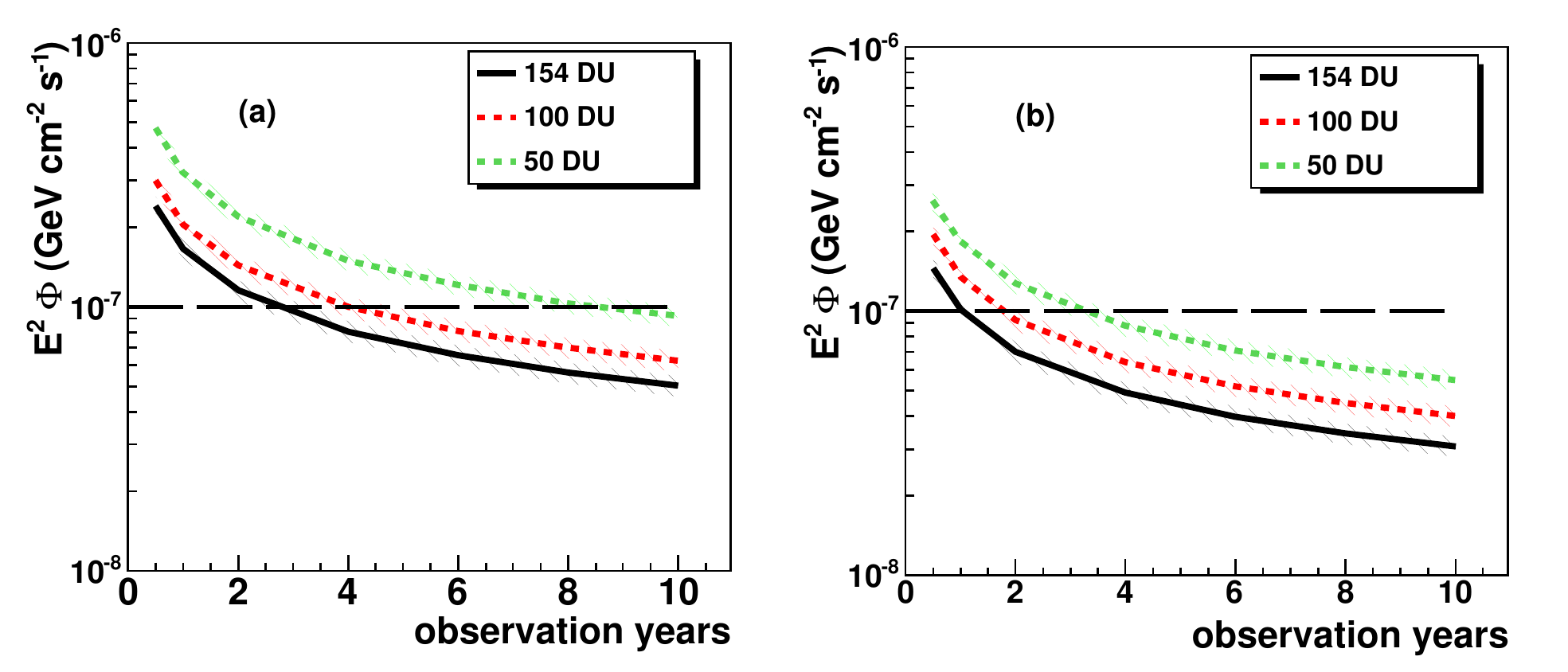}
\caption{
Discovery fluxes as a function of the observation years for a $5\sigma$ C.L., $50\%$
probability (a) and $3\sigma$ C.L., $50\%$ probability (b) 
for a $E^{-2}$ neutrino spectrum with a $100\,$TeV cutoff for different
detector sizes (see text). The bands represent the variation due to
the uncertainty on the normalisation factor of the conventional Bartol neutrino
flux. The long-dashed lines indicate the predicted neutrino flux estimated in Sec.~4.1.}
\label{fig:DiscFlux_DU}
\end{figure*}

Recently an analysis has been performed on 43 months of Fermi data in the energy
range $20\,\mathrm{GeV}<E_{\gamma}<300\,$GeV \cite{Weniger}. This analysis
covered several regions near the Galactic centre, some of which overlap with the
bubble regions. This revealed the presence of an excess in the gamma ray flux
around $130\,$GeV. This peak has been interpreted as an indication of dark
matter annihilation into two photons. An alternative explanation, put forward in
a separate analysis of the same data (in the range
$80\,\mathrm{GeV}<E_{\gamma}<200\,$GeV) \cite{Profumo}, is that the excess can
be interpreted as a steepening of the power law spectrum of the Fermi bubble
gamma rays at around $130\,$GeV. Such a steepening has a significant effect on
the results of the present analysis. Therefore, a Fermi bubble neutrino spectrum
that follows a power law spectrum according to $d\Phi^{\infty}_{\nu}/dE$ with
$K_{0\nu}(E) = 10^{-7}\,\mathrm{GeV\,cm^{-2}\,s^{-1}}$ up to $130\,$GeV and
extends to higher energies proportional to $E^{-2.3}$, has been investigated. In
this case the resulting live time required for discovery is 6 years, which is
well in the expected lifetime of the KM3NeT detector.

If no statistically significant excess of neutrino events will be found, upper limits can
be set for specific neutrino emission models. In Fig.~\ref{fig:SensFlux_km3net} the 
average upper flux limits at $90\%$ of C. L. (computed following the Feldman and 
Cousins prescription \cite{Feldman}) are reported for the three neutrino spectra 
investigated.  If the observed gamma flux is of purely hadronic origin and the source
is fully transparent for gamma rays, the expected neutrino spectrum corresponds 
to $K_{0\nu} = 1\times10^{-7}\,\mathrm{GeV\,cm^{-2}\,s^{-1}}$ (see Sect. 4.1). If the  
gamma emission is not fully hadronic and  the hadronic and non-hadronic emission 
mechanisms yield the same spectral shapes, upper limits on the percentage of 
hadronic emission can be extracted from the upper flux limits in Fig.~\ref{fig:SensFlux_km3net}
(see right vertical scale).

Since neutrino telescopes have a modular design, the detector concept allows for
a staged implementation with continuously increasing science capabilities. 
Therefore, the capability of the telescope to detect neutrinos from the Fermi
bubbles has been explored as a function of an increasing number of DUs. 
Detector geometries composed of 50, 100 and 154 DUs with the same average 
DU distance ($180\,$m) have been investigated (see Fig.~\ref{fig:Footprint}). 
These geometries provide increasing
active volumes of 1, 2.1 and $3.2\,\mathrm{km}^3$ for 50, 100 and 154 DUs, respectively. 
The discovery fluxes have been calculated and are shown in Fig.~\ref{fig:DiscFlux_DU} for $5\sigma$ C.L.,
$50\%$ probability (a) and $3\sigma$ C.L., $50\%$ probability (b).  
The number of years needed for the discovery does not decrease
linearly with the increasing number of DUs, being about 8 years for 50 DUs, 4
years for 100 DUs and 2.5 years for 154 DUs. A neutrino spectrum with
normalisation $ K_{0\nu} \approx 1 \times10^{-7}\,\mathrm{GeV\,cm^{-2}\,s^{-1}}$ 
with a $100\,$TeV cutoff energy is assumed here.


\section{Conclusions}
The discovery of an intense gamma-ray flux from two large areas around the
Galactic centre, the Fermi bubbles, has stimulated estimates of the capability
of neutrino telescopes to discover neutrinos from this region
\cite{Lunardini,Cholis}. The Fermi bubbles extend over a large part of the sky
and have a significant intensity in gamma rays. Assuming these gamma rays are of
hadronic origin and the spectrum extends to the multi-TeV range, this analysis
shows that high energy neutrinos are expected from near
the Galactic centre. Telescopes located in the Mediterranean Sea, such as
ANTARES and KM3NeT, have a large visibility of the region around the Galactic
centre. Due to its multi-km$^3$ scale, the future KM3NeT telescope is the ideal
instrument to observe neutrinos from the Fermi bubbles. Note that, due to its
location, IceCube has a considerably reduced visibility of the bubbles.

In this paper we present an analysis addressing the sensitivity of KM3NeT to a
neutrino flux from the Fermi bubbles that is consistent with the measured gamma
ray flux, assuming that the latter fully originates from hadronic processes. 
Based on a complete chain of Monte Carlo simulations, the time of KM3NeT
operation that is required to detect this neutrino flux at a significance of
$5\sigma$ with $50\%$ probability has been estimated. The result depends
strongly on the shape of the neutrino energy spectrum assumed. In the case of an
$E_\nu^{-2}$ spectrum with an exponential cutoff at $100\,$TeV, we expect to
obtain a $3\sigma$-evidence in a few months and to claim a discovery after about
1\,year of data taking. The non-observation of a signal would severely constrain
models of neutrino production through hadronic acceleration processes in the
Fermi bubbles.

\section*{Acknowledgements}
The research leading to these results has received funding from the European
Community's Sixth Framework Programme under contract n$^\circ$ 011937 and the
Seventh Framework Programme under grant agreement n$^\circ$ 212525.

\label{}








{\raggedright

}
\end{document}